# Improving VANET's Performance by Incorporated Fog-Cloud Layer (FCL)

Ghassan Samara
Computer Science Department
Zarqa University
Zarqa- Jordan
gsamarah@zu.edu.jo

Mohammed Rasmi
Departmint of Internet Technology
Zarqa University
Zarqa- Jordan
mmousa@zu.edu.jo

Nael A. Sweerky
Computer Science Department
Zarqa University
Zarqa- Jordan
sweerky_nael@hotmail.com

Essam Al daoud
Software Engineering Department
Zarqa University
Zarqa- Jordan
essamdz@zu.edu.jo

Amer Abu Salem
Computer Science Department
Zarqa University
Zarqa- Jordan
abusalem@zu.edu.jo

*Abstract*— Because of its usefulness in various fields including as safety applications, traffic control applications, and entertainment applications, VANET is an essential topic that is now being investigated intensively. VANET confronts numerous challenges in terms of reaction time, storage capacity, and reliability, particularly in real-time applications. As a result, merging cloud computing and cloud computing has recently been researched. The goal of this study is to develop a system that merges the fog and cloud layers into a single layer known as the included fog-cloud layer. To lower the time it takes for real-time applications on VANETs to respond while also improving data flow management over the Internet and achieving an efficient perception service while avoiding the high cost of cloud connectivity.

## I. INTRODUCTION

The number of vehicles on the road has increased over the last two decades, resulting in increased traffic congestion and crashes, as well as increased emissions from fuel use. Many individuals have died as a result of these issues, as well as the devastation caused by global warming. To address these problems, the VANET emerged as a beneficial option for making road situations safer and more comfortable for drivers, passengers, and other persons [1].

Vehicles can communicate with one another via the VANET by sending and receiving messages. The VANET includes a variety of services such as traffic control services to determine the best route and avoid traffic congestion, safety services that must send real-time messages to prevent critical circumstances such as accidents, and entertainment services that make long-distance travel more enjoyable for passengers. [2, 3, 4, 5, 6, 7, 8, 9, 10, 11] .

Vehicles can now be supplied with a variety of wireless sensors, storage, wireless connection modules, and computer services thanks to VANET advances. Massive volumes of data have been generated as the number of these units mounted on vehicles has increased, as has the number of vehicles [12]. Some applications, like as self-driving apps, necessitate extensive and complicated processing capabilities as well as a huge storage capacity. As a result, VANET faces significant hurdles in extending and achieving network requirements to meet users' data needs [13, 14, 15].

The design of a VANET is known to be static. Due to an imbalance in data flow caused by traffic imbalance and the position of the node on the network, it is difficult to plan and spread out [16, 17]. Because the nodes are widely distributed in geographically distant places, it is difficult to regulate VANET because of its extremely dynamic structure and unreliable connections [18].

Cloud computing in a VANET conducts huge data processing in a short period of time and has a high storage capacity, but it confronts significant obstacles due to the VANET's quick development and pace of growth, as well as the high demand for data inside it [19, 20, 21]. The two most significant difficulties that cloud computing cannot address are data flow speed and response time. As a result, an intermediary layer dubbed the fog layer should have existed to overcome these issues and operate as a lite form of cloud computing [22, 23].

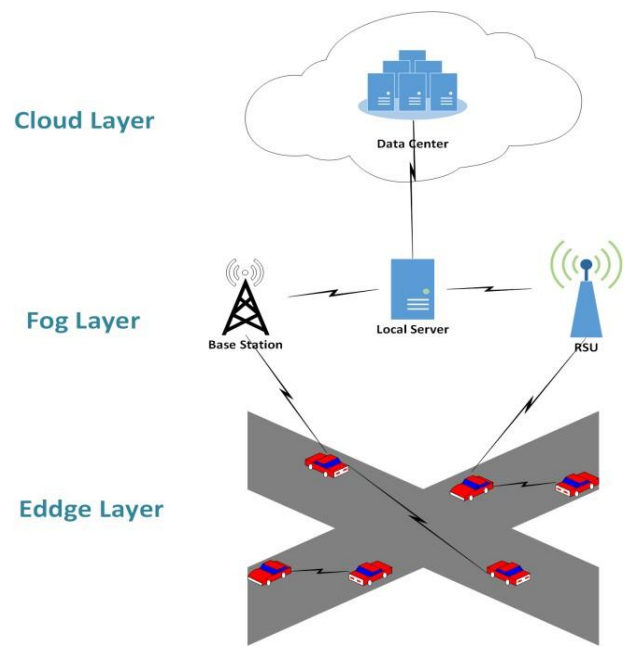

Fig. 1. The three-layer model SDN-VANETs

The three-layer approach (edge layer, fog layer, and cloud layer) depicted in Figure 1 is the perfect and clever method for dealing with the nature of VANETs in terms of extremely dynamic structure, high bandwidth, unstable connection, and high temporal sensitivity [24, 25]. Each layer in this paradigm serves a variety of purposes while ultimately complementing the others. On the three-layer paradigm, data is processed to a certain extent in the edge layer. When this limit surpasses this layer's capabilities, these jobs are changed in the fog layer. When the processing of a task surpasses the capacity of the fog layer, it is routed to the cloud layer [26, 27, 28].

Although the three-layer approach has solved many difficulties, it confronts a number of obstacles in regulating the management of network resources and communications because it is a decentralized paradigm. As a result, the best answer to these problems is to employ a software-defined network (SDN). The goal of SDN is to isolate the control level from the data level, allowing the VANET's control level to be centralized via SDN [29, 30]. As a result, all nodes on the network are monitored, which improves control and data flow. This also supports optimal network resource utilization and boosts network flexibility, efficiency, adaptability, and scalability [31, 32, 33].

One of the most significant issues that VANETs face is security. Despite the importance of security, there has been little research in this field. The information contained within VANET must be safeguarded, and attackers must be stopped from altering it. The system should be able to track the driver's duty while safeguarding the privacy of both the driver and the passengers [34, 35]. There are numerous types of VANET attacks, including vehicle illegal tracking, exposing their identification, and denial of service assaults. This critical component is safeguarded by preventing VANET assaults by integrating authentication, verification, and the usage of a public key to encrypt data [36, 37].

Despite advancements, the (edge-fog-cloud) model still suffers from latency and data flow issues. The suggested solution aims to construct a typical example that enhances reaction speed and data flow speed, particularly in real-time applications, by merging the cloud layer and the fog layer, and the new merged layer is referred to as the fog-Cloud layer (FCL). As a result, it minimizes the time required to send data from remote data centers in the cloud layer to edge layer nodes. Data that requires a lot of processing capacity is processed in fog-cloud data centers (FCDC) in FCL, which increases the overall performance of the service implementation.

## II. LITERATURE REVIEW

The authors of presented a system based on fuzzy logic in [38]. They presented two intelligent systems that were utilized to choose the right layer to make the choice and process data from automobiles in the edge layer, delivering significant efficiency in latency reduction. A variety of parameters are considered, including (Data Size, Time Sensitivity, Vehicle Relative Speed with Neighboring Vehicles, and The Number of Neighboring Vehicles). Messages on beacons

Data are obtained from nearby vehicles, such as direction, velocity, and current position, are beneficial to the proposed system's choice.

[39] presents a case study of 5G and mobile edge computing to deliver real-time context-aware connectivity. As a result, VANET is expected to know the attributes and location of any node, as well as the circumference of that node. In this system, new fifth-generation communication architectures are merged with mobile edge computing principles to build a shared system capable of handling real-time applications while successfully dealing with delay and low latency. The advantages of mobile edge computing are leveraged due to its proximity to network users, as well as the advantages of the fifth-generation network in terms of speed, which presents enormous possibilities when these two technologies are combined.

The authors of [40] proposed a concept (fog vehicle computing) that is based on collaboration between cars and the end-user edge. All computation and communication is handled by a VFC. The model exploits and employs the resources of cars in the edge layer, such as memory and processors, to allow these vehicles to communicate and conduct data processing activities. Vehicles are chosen depending on certain characteristics. Vehicle parking patterns are monitored and their capabilities are maximized, considerably improving the quality of service offered inside the network.

The authors of [41] created the VANET-Cloud vehicle cloud computing paradigm. The model is divided into two subsystems: the permanent cloud and the temporary cloud. The permanent cloud is built on a standard cloud system that offers consistent services to users at the edge layer. The second type is temporary clouds, which are made up of a group of vehicles that work together to form a temporary cloud that provides services to users in the edge layer, thereby expanding cloud layer coverage, increasing processing capacity, and better meeting user needs. The system delivers low-cost services and improves road safety by collecting information from the edge layer, evaluating it, and delivering it to vehicles on the road so that they can take appropriate action in the event of an emergency. Drivers can also earn money by making use of the computing resources built in their vehicles. This model extends the cloud layer, which contains fixed stations, to cars in the edge layer.

The authors of [42] introduced an architecture that combined fog computing and software-defined networking (SDN) and applied a service called Cooperative Service in Vehicular Fog Computing (CS-VFC), which aims to improve the service provided by the fog layer and increase the effectiveness of the bandwidth through the integration of the service in the cloud layer and the fog layer and improve the coordination between the fog layer and the cloud layer. The concept employs a scheduling algorithm that makes decisions within a software-defined networking controller, allowing for easier decision-making and faster data flow.

The authors of [43] divided the cloud into three interactive layers. The first tier is known as the (vehicle cloud), and it consists of a collection of cars within the network's scope. The second layer is known as the roadside unit cloud, and the third layer is known as the central cloud, which comprises data centers connected to the Internet. Based on a resource reservation mechanism, it proposed a method for optimally utilizing various cloud resources in the event of vehicle mobility. This system combines the redundant resources in intelligent transportation systems (cars, local data centers, and roadside devices) to form a massive cloud resource that aids in the provision of services to vehicles. A method has been developed to manage and use these cloud resources efficiently, as well as to establish a solid plan to cope with the mobility of vehicles whose resources are being used by the system.

The authors of [44] presented some concepts about fog computing and cloud computing scenarios, stressing many issues that these scenarios faced when they were researched and evaluated in terms of latency. This model proposes two ways for executing distributed services in fog and cloud computing. The first approach is known as Sequential Service Execution. This technique gradually splits service execution into dedicated resources, and each service is done in one of the resources chosen to best meet the capacity of that resource and the needs of the vehicle. The other option is known as parallel service execution, and it separates the service's execution into assigned resources in parallel. Both techniques are constantly seeking for the best workflow for service execution and resource utilization.

In [45], the authors propose a mobile computing approach in which the user uses local cloudest to them instead of incurring delay losses, and in the absence of nearby cloudest, the user connects to the main cloud, lowering the delay as much as feasible. They proposed cloudest, a collection of servers with high processing power and big capacity. It is linked to the Internet and is situated in the fog layer. Because of its proximity to the edge layer, Cloudest lowers the time it takes for data to travel from the cloud layer to the edge layer. This paradigm combines the benefits of fog computing and cloud computing and is used in network development and service delivery with high quality and speed, producing good outcomes when compared to the old system.

## III. INCORPORATED FOG-CLOUD LAYER

Because the fog layer is close to the edge layer, its response time is quite short when compared to the cloud layer. However, it has limited storage and processing capabilities. The cloud layer is distinguished by its high data processing capability and vast storage capacity, but it is hampered by its geographical distance from the edge layer, resulting in a delayed response, particularly in real-time applications that require a fast response time. The larger the distance between the cloud layer and the edge layer, the poorer the performance, efficiency, security, and dependability [46, 47].

The proposed system combines the advantages of the fog layer with the cloud layer, so that each layer complements and compensates for the faults of the other. Bringing together resources

Fog layer and cloud layer combine to form massive and powerful fog-cloud computing that serves clients at the edge layer. Because of its proximity to the edge layer, combining the two levels into one layer increases response time while simultaneously maximizing processing and storage capacity. Because there is no need to process data in geographical regions remote from the edge layer in this paradigm, a data center named (FCDC) is positioned in the FCL layer.

Figure 2 depicts the fog-cloud layer, which is made up of FCDCs spread across different geographical locations. This distribution is determined by customer needs, service demand, and transportation congestion in these areas. FCDCs deliver a variety of services with low latency, satisfying users at the edge layer. All activities and requests are handled, and the same services as provided by the cloud layer are provided, but at a faster and more efficient rate.

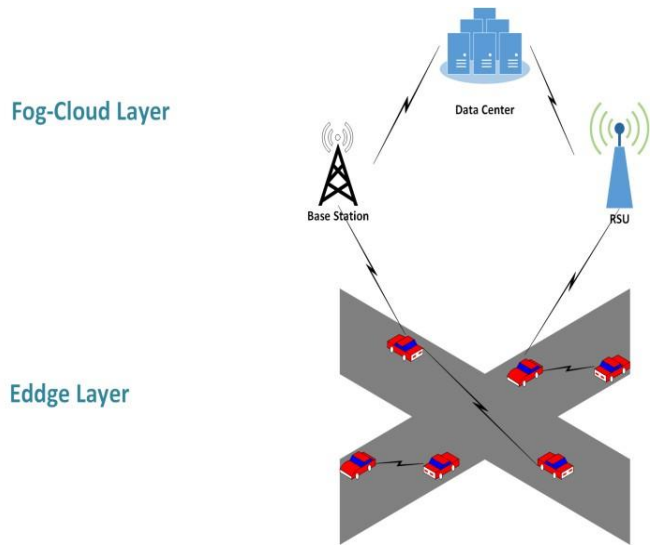

Fig. 2. Fog-Cloud Layer and Edge Layer

Data is received from vehicles located throughout a vast geographical region and under various conditions in Fog-Cloud Computing. FCL evaluates and analyzes this data, and an immediate decision is made and transmitted to all vehicles in this region, preventing numerous accidents from occurring if this decision is delayed.

During vehicle movement, VANETs use RSU as a gateway to communicate with the fog-cloud layer. Cars employ the fog-cloud layer FCDCs to collect the essential information that assists them while driving, so that the FCDCs have complete knowledge about the vehicles and their geographical location. The edge layer gathers information about the cars through sensors mounted in the vehicles. Vehicles that require safety messages are determined based on the traffic situation, so that each vehicle is in the proper position to deal with crises.

Connection is a critical issue in the VANET, and it is seen as a major issue, particularly in applications that require high connectivity. For example, the connectivity with the rest of the VANTs resources must be available in the self-drive automobile. The proposed model has high connection, allowing various applications to run smoothly and without interruption in order to avert disasters and the rest of the cars. Aside from the low latency provided by this approach because to its proximity to FCDCs.

Table (I) shows the differences between the fog-cloud layer and the cloud layer. The fog-cloud layer scales better than the cloud layer. While the fog-cloud layer can handle real-time applications, the cloud layer cannot because its latency is too high in comparison to the fog-cloud layer. When data from the edge layer arrives in real-time with low latency and is analyzed, it provides an accurate report regarding road traffic. The fog-cloud layer's network architecture is decentralized, whereas the cloud layer's network design is centralized. The fog-cloud layer supports mobility within VANETs, whereas the cloud layer does not. The fog-cloud layer collects and analyzes information about cars while they are going and in various environmental conditions, whereas the cloud layer is unable to do so. The data center is located near the edge layer, but the data center on the cloud layer is located far away from the edge layer. While the cloud layer is unaware of the position of each car linked to the network, the fog cloud layer is.

TABLE I.   COMPARATIVE BETWEEN FCL AND CLOUD LAYER

|  | Fog-Cloud Layer | Cloud Layer |
|---|---|---|
| Scalability | High | Average |
| Mobility support | Yes | Limited |
| Real-Time | Yes | No |
| Low-latency | Yes | No |
| Network Architecture | Decentralized | Centralized |
| Location wareness | Yes | No |
| Data Center Location | Near | Far |

## IV. CONCLUSION

This research offered a model that merges the cloud and fog layers into a single layer called (The Fog- cloud layer). The comparison of the fog-cloud layer and the cloud layer was made from different perspectives, and the fog-cloud layer was preferred in several areas, including scalability, mobility support, real-time, low-latency, and network architecture.

The proposed approach reduces response time for real-time applications while simultaneously improving data flow management.

Because of its proximity to the FCDCs, the Internet. As a result, the cost of connecting to FCDC is reduced.

The proposed model's security and privacy issues, as well as the validation and verification procedure, are critical. In the future, we hope to increase the model's security and privacy..